\RequirePackage{fix-cm}
\documentclass[smallextended]{svjour3}       
\smartqed  
\usepackage{appendix}
\usepackage{amsmath}
\usepackage{graphicx}
\usepackage{lineno}
\usepackage{array}
\usepackage{longtable}
\newcommand*\patchAmsMathEnvironmentForLineno[1]{%
\expandafter\let\csname old#1\expandafter\endcsname\csname #1\endcsname
\expandafter\let\csname oldend#1\expandafter\endcsname\csname end#1\endcsname
\renewenvironment{#1}%
{\linenomath\csname old#1\endcsname}%
{\csname oldend#1\endcsname\endlinenomath}}%
\newcommand*\patchBothAmsMathEnvironmentsForLineno[1]{%
\patchAmsMathEnvironmentForLineno{#1}%
\patchAmsMathEnvironmentForLineno{#1*}}%
\AtBeginDocument{%
\patchBothAmsMathEnvironmentsForLineno{equation}%
\patchBothAmsMathEnvironmentsForLineno{align}%
\patchBothAmsMathEnvironmentsForLineno{flalign}%
\patchBothAmsMathEnvironmentsForLineno{alignat}%
\patchBothAmsMathEnvironmentsForLineno{gather}%
\patchBothAmsMathEnvironmentsForLineno{multline}%
}

\usepackage{amssymb,amsmath,graphicx,color,amsfonts}
\usepackage{bm}
\usepackage[tight]{subfigure}
\usepackage[export]{adjustbox}
\usepackage{braket}
\usepackage[colorlinks=true,citecolor=blue,linkcolor=blue]{hyperref}
\usepackage[table]{xcolor}
\usepackage{cancel}
\usepackage{multirow}
\usepackage{fixltx2e}
\usepackage{blindtext}

\newcounter{oldtocdepth}

\newcommand\beq{\begin{equation}}
\newcommand\eeq{\end{equation}}
\newcommand\bea{\begin{eqnarray}}
\newcommand\eea{\end{eqnarray}}

\begin{document}

\title{
Towards Quantum Simulating Non-Abelian Gauge Theories}

\author{Indrakshi Raychowdhury}

\institute{ Maryland Center for Fundamental Physics and Department of Physics,\\ 
University of Maryland, College Park, MD 20742, USA
              \\\email{iraychow@umd.edu}      
}

\date{Received: DD Month YEAR / Accepted: DD Month YEAR}

\maketitle



\begin{abstract}
In this article, we review some of the recent developments towards the future goal of quantum computing or quantum simulating lattice QCD. This includes a novel theoretical framework developed for non-Abelian gauge theories that is the first necessary step towards this goal. We also review some immediate application of this framework in context of both digital and analog quantum simulation of SU(2) lattice gauge theory coupled with staggered fermions.
\end{abstract}


\section{Introduction}
\label{sec:intro}
\noindent

Gauge field theories lie at the core of understanding the fundamental interactions of nature including electromagnetic, weak and strong interactions \cite{quigg2013gauge}. The standard model of particle physics is described by a non-Abelian gauge theory. Quantum Chromodynamics (QCD) describes the interaction of quarks mediated via gluons \cite{MARCIANO1978137}, which are the quanta of gauge fields governed by local symmetry group SU(3). Gauge theories formulated on space-time lattice  \cite{Wilson:1974sk} (lattice QCD) have proved to be an extremely useful non-perturbative technique that uses classical computational resources at its fullest extent for the past four decades \cite{CREUTZ1983201,joo2019status}. However, albeit using world's largest supercomputers, there exists certain inaccessible regimes for lattice-QCD, mostly due to presence of the infamous sign problem \cite{de2010simulating}. One must note that, these obstacles are not due to features of QCD, but are due to the conventional lattice-QCD calculational techniques. This calls for drastically new computational techniques, supplemented by  ideas emanating from less trodden paths in gauge theory formulation. Following Feynman \cite{feynman1982simulating}, the first shot towards this would be to try to understand QCD by quantum simulation.

Currently, we are witnessing a change of paradigm in computational technique with the ongoing efforts and developments towards building the universal quantum computers \cite{arute2019quantum}. It is also exciting that the quantum computing resources are becoming available quite rapidly \cite{castelvecchi2017ibm}. In this scenario, a vast community of researchers engaged in developing quantum computation hardware are also keen to learn the probable use of quantum computers as well as resource estimation for the same. At the same time, researchers across several field are putting great effort to identify the problem that would actually benefit from the \textit{quantum advantage} followed by the effort to formulate the problem in a framework suitable for quantum computers and then developing quantum algorithms. Also, for the initial era (Noisy Intermediate Scale Quantum or NISQ) \cite{preskill2018quantum} it is important to construct problem specific analog quantum simulators \cite{buluta2009quantum} for certain problems that would be immediately more useful than the more ambitious digital quantum computer.

As mentioned before, for the  lattice gauge theory community, quantum computation offers hope to go beyond the capability of conventional lattice-QCD calculations. At the same time, quantum computation for lattice QCD requires a lot of ground works as the theoretical framework, problem to address and the calculational techniques to be used in the upcoming quantum computing era, are significantly different from what has been used in conventional lattice-QCD computations so far. To elaborate on this, Hamiltonian framework  is the natural setting for quantum computation unlike the Euclidean path integral formalism used so far extensively. While, most of the efforts in lattice QCD was to understand different static properties of the theory, the upcoming quantum computation mainly would focus to understand the dynamics of the theory that is natural with Hamiltonian simulation. Also, lattice QCD calculations are performed by Monte-Carlo simulation that uses important sampling of gauge field configurations; whereas Hamiltonian simulation involves a completely different set of methods which include construction of Hilbert space, imposing constraints, diagonalization of Hamiltonian matrix and constructing unitary operators for time evolution.

The past decade witnessed highly active interdisciplinary efforts towards quantum simulating lattice gauge theories  \cite{Banuls:2019bmf}. Starting with the first few proposals for analog quantum simulation \cite{Zohar:2012xf,Banerjee:2012pg,Zohar:2013zla,Banerjee:2012xg,Zohar:2014qma,Zohar:2015hwa}, there has been considerable progress in this field. However, quantum simulating the full-fledged QCD is still a very far fetched idea, while researchers across the globe are working to analyze simpler yet qualitatively similar theories. A strong candidate of such simple theory that shares many qualitative feature of QCD is Quantum electrodynamics (QED) that is an Abelian gauge theory with gauge group U(1). Even a simpler form of QED, that is the Schwinger model (QED in $1+1$ dimensions) has been particularly popular in this context \cite{Banuls:2019bmf,kasper2016schwinger,Davoudi:2019bhy,klco2018quantum,shaw2020quantum,stryker2019oracles,chakraborty2020digital}. There exists over hundreds of articles in the literature on the topic of the quantum simulation of the Schwinger model including at least 3 actual experimental demonstration of analog quantum simulation of the same \cite{martinez2016real,Mil:2019pbt,Yang:2020yer,} and one rigorous quantum algorithm for simulating Schwinger model on a universal quantum computer \cite{shaw2020quantum}. Works are also in progress by different teams of scientists towards a more efficient and large scale quantum simulation of Schwinger model. Other simpler gauge theories that provide a useful test-bed of novel quantum simulation schemes are the discrete gauge theories such as $\mathbb Z_2$ gauge theories. There has been experimental demonstration of quantum simulating $\mathbb Z_2$ gauge theories on a small lattice \cite{schweizer2019floquet,gorg2019realization}.

In contrast to all of these developments, there is a very little progress towards quantum simulating any non-Abelian gauge theory, even for the simpler gauge group SU(2). Few early  proposals for analog simulation of SU(2) theory is there \cite{Banerjee:2012xg,tagliacozzo2013simulation,Zohar:2012xf}. But none of these have been experimentally implemented so far. It is only recently, when there has been a community wide active interest in generalizing the Schwinger model proposals and/or algorithms for a non-Abelian gauge group such as SU(2) \cite{davoudi2020towards,Zohar:2019ygc,Zohar:2018cwb}. 

The major reason behind the lack of progress with quantum simulation of non-Abelian gauge theory is the non-Abelian gauge redundancy and one needs to extract out the physical Hilbert space of the theory for any practical computation. Otherwise, the noisy quantum computer will actually be lost into the unphysical degrees of freedom of the system and it will be practically impossible to extract out any physical result. With the conventional Hamiltonian framework of non-Abelian gauge theory, imposing the constraints and extraction of the physical Hilbert space is quite a non-trivial job \cite{davoudi2020towards}. Also, the Hamiltonian acts on different irreducible representations of the non-Abelian gauge group differently. Hence, implementing the same in terms of quantum circuits or simulating by a quantum simulator is possible in principle, but is rather unnatural to do so.

In this review, we briefly discuss the theoretical framework for Hamiltonian simulation of SU(2) gauge theory in section \ref{sec:toolbox} including a promising and alternate reformulation of the original framework that is expected to be useful in the context of quantum simulation. Next, we briefly mention some of the applications of this novel framework in context of digital as well as analog quantum simulation of SU(2) gauge theory in section \ref{sec:app}.

\section{The theoretical framework}
\label{sec:toolbox}
\noindent

The natural framework for quantum computation is the Hamiltonian framework. Right after Wilson proposed the lattice regularized version of gauge theory, the Hamiltonian formalism of the same was developed by Kogut and Susskind \cite{Kogut:1974ag}. In the next subsection, we will highlight the key features of the Kogut-Susskind formulation of lattice gauge theory and then will also review a reformulation of the same, in a form more suitable for practical calculation.

\subsection{Kogut-Susskind Hamiltonian formalism}
\label{sec:KS}
\noindent
\begin{figure}
\centering
 \includegraphics[width=0.8\textwidth]{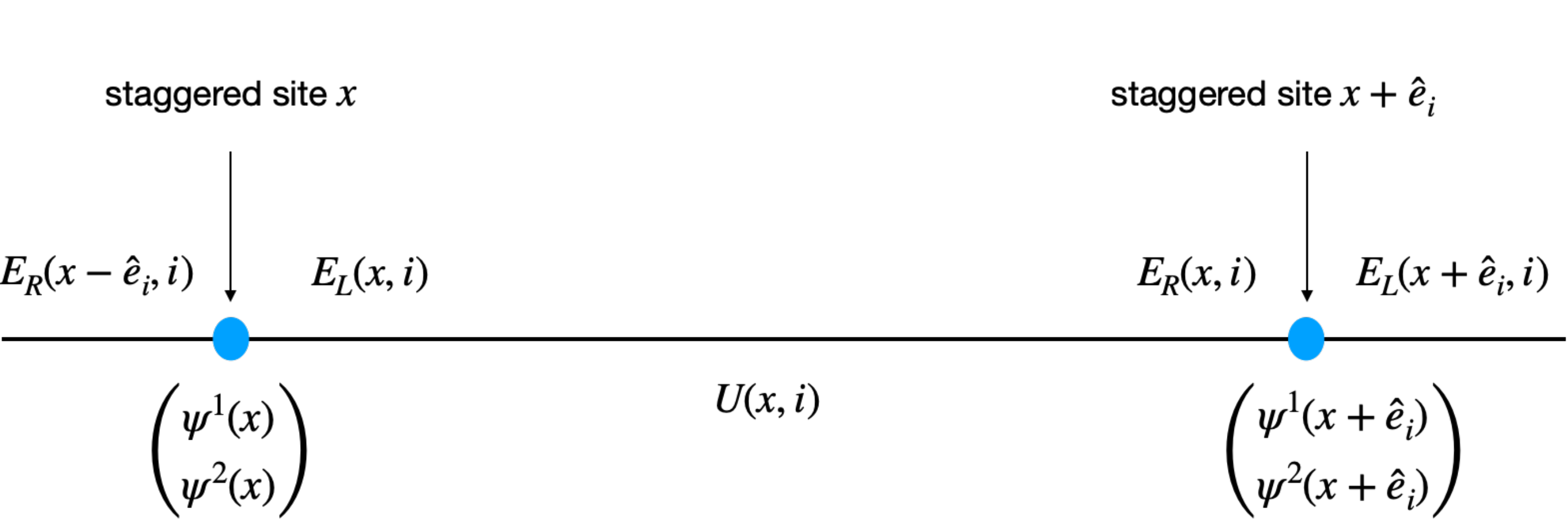}
\caption{Electric fields, link operator or holonomy and matter fields defined on a link of a spatial lattice that starts from site $x$ along direction $i$.}
\label{fig1}       
\end{figure}
The Kogut-Susskind (KS) Hamiltonian describing SU(2) Yang Mills theory coupled to staggered fermions on $(d+1)$-d (d dimensional spatial lattice and continuous time) \cite{Kogut:1974ag} can be written as:
\begin{eqnarray}
H&=& \frac{g^2a}{2} \sum_{x}\sum_{i=1}^{d} \sum_{a=1}^3{E^a}(x,i){E^a}(x,i) \nonumber \\
&& + m\sum_{x}(-1)^x\left[\psi^{\dagger}(x)\cdot \psi(x) \right]  \\
&& + \frac{1}{2a}\sum_{x}\sum_{i=1}^{d}\left[\psi^{\dagger}(x) U(x,i) \psi(x+\hat e_i)+{\rm h.c.} \right]\nonumber \\
&& + \frac{4}{ag^2} \sum_{x_{ij}} \left[\mathrm{Tr}\, \left[U(x,1)U(x+\hat e_i,j)U^\dagger(x+\hat e_i+\hat e_j,i) U^\dagger(x+\hat e_j,j)\right] + h.c. \right] \nonumber
\label{eq:HKS}
\end{eqnarray}
Where, $x$ is a lattice site and $x_{ij}$ is a plaquette in $i-j$ plane with one corner at site $x$. The color electric fields $E_{L/R}^{a}$ are defined at the left $L$ and right $R$ sides of each link as shown in Fig. \ref{fig1} and they satisfy the following commutation relations (su(2) algebra) at each end:
\begin{eqnarray}
[E_L^a(x,i),E_L^b(x',i')]&=&i\epsilon^{abc}\delta_{xx'}\delta_{ii'} E_L^c(x,i),
\nonumber\\
{[E_R^a(x,i),E_R^b(x',i')]}&=&i\epsilon^{abc} \delta_{xx'}\delta_{ii'} E_R^c(x,i),
\nonumber\\
{[E_L^a(x,i),E_R^b(x',i')]}&=&0,
\label{eq:ERELcomm}
\end{eqnarray}
where $\epsilon^{abc}$ is the Levi-Civita symbol.  The electric fields and the gauge link satisfy the following quantization conditions at each site,
\begin{eqnarray}
[E_L^a(x,i),U(x',i')]=-\frac{\sigma^a}{2}\delta_{xx'}\delta_{ii'}U(x,i),
\nonumber\\
{[E_R^a(x,i),U(x',i')]}= U(x,i)\delta_{xx'}\delta_{ii'}\frac{\sigma^a}{2},
\label{eq:EUcomm}
\end{eqnarray}
where $\sigma^a$ are the Pauli matrices\footnote{ For general SU(N) with $N>2$, these will be the generalized Gell-Mann matrices $\lambda^a$, generating the Lie algebra associated to the special unitary group SU(N). For the specific case of SU(3), the $3\times 3$ standard Gell-Mann matrices would play the same role}. 
The Hamiltonian in (\ref{eq:HKS}) is gauge invariant as it commutes with the  Gauss' law operator,
\begin{equation}
G^a(x)=\sum_{i=1}^{d}E^a_L(x,i)+E^a_R(x,\bar i)+\psi^\dagger(x) \frac{\sigma^a}{2} \psi(x)
\label{eq:Ga}
\end{equation}
at each site $x$.  The physical sector of the Hilbert space corresponds to the space consisting of states annihilated by (\ref{eq:Ga}). 

Also note that, the left and right electric fields on a particular link $(x,i)$ are parallel transported to each other along the link and hence the electric part of the Hamiltonian (\ref{eq:HKS}) becomes,
\bea
\label{eq:AGL_KS}
\sum_{a=1}^3{E^a}(x,i){E^a}(x,i)=\sum_{a=1}^3{E_L^a}(x,i){E_L^a}(x,i)=\sum_{a=1}^3{E_R^a}(x,i){E_R^a}(x,i).
\eea
The SU(2) gauge theory Hilbert space can be mapped to a space of rigid rotors for each link and can be characterized by states $|j,m_L,m_R\rangle$, where $j=j_L=j_R$. The link operator acting on these states yield a linear combination of irreducible representations (irreps) with $j\rightarrow j'=j\pm \frac{1}{2}$ along with the corresponding Clebsch Gordan coefficients. 

Extraction of the physical Hilbert space from the space of rigid rotors is non-trivial. First, one needs to solve the Gauss' law constraint at each site of the lattice to get the loop variables of the theory. However the proliferation of loop space make it an over-complete basis. Hence it's necessary to solve a further set of constraints, namely the Mandelstam constraint on the lattice. For higher dimension and large lattices, solving Mandelstam constraints is an almost impossible task. 

In summary, working with physical Hilbert space built out of the space of rigid rotors is extremely difficult and engineering the same in an analog experiment or in a quantum algorithm is one of the greatest challenges in the field. The next level hurdle in the whole program is the engineering of the action of the Hamiltonian in this basis.

In the following, we review a reformulation of this original Kogut-Susskind formalism of Hamiltonian lattice gauge theory towards the goal of quantum simulating the same. 


\subsection{Prepotential formalism}
\label{sec:pp}
\noindent

\begin{figure}
\centering
 \includegraphics[width=0.8\textwidth]{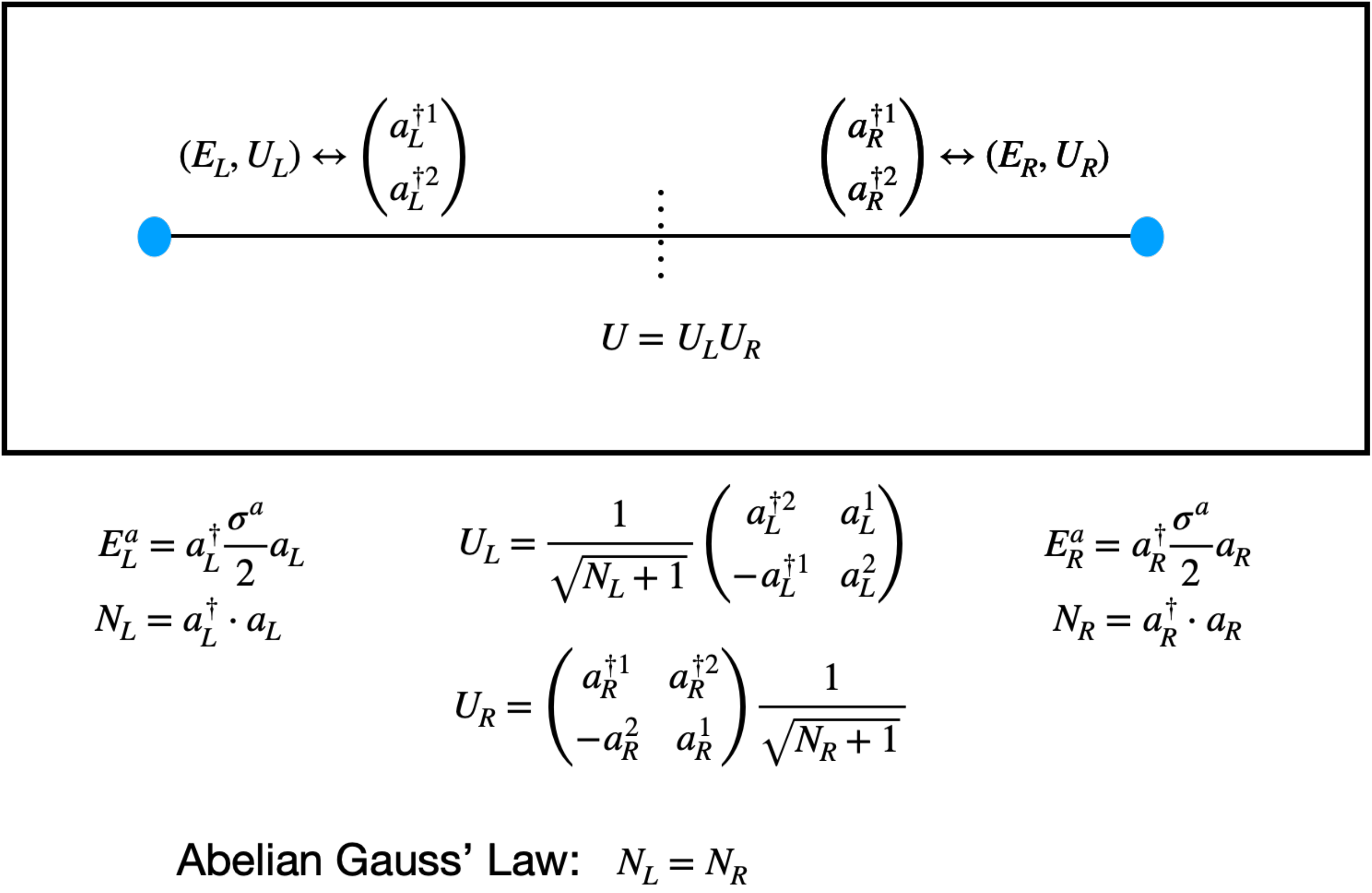}
\caption{Within prepotential formulation of pure SU(2) gauge theory, the canonical conjugate variables are reconstructed in terms a set of independent harmonic oscillator doublets at both ends of a particular link. The link operator also breaks into a left matrix and a right matrix, each constructed out of left and right oscillators respectively. The only connection between the left and the right end of the link exists via an Abelian constraint that equates the total number of oscillators at the left end to that at the right end. This Abelian constraint is a consequence of (\ref{eq:AGL_KS})}.
\label{fig:pp}       
\end{figure}

Prepotential formulation of pure lattice gauge theory has been developed and refined over the past two decades \cite{Mathur:2004kr,Mathur:2007nu,Mathur:2010wc,Anishetty:2009ai,Anishetty:2009nh,Anishetty:2014tta,Raychowdhury:2013rwa,Raychowdhury:2014eta,Raychowdhury:2018tfj,mathur2010n}. Prepotential operators are basically Harmonic oscillator doublets that are defined at each end of a link as shown in the Fig. \ref{fig:pp}. One can then rewrite  the canonical conjugate variables in terms of the oscillators or prepotentials to satisfy the original set of commutation relations and hence one can work solely with the prepotential operators forgetting about the original variables as well as Hilbert space of the theory. 

The most striking feature of the prepotential formalism is that the prepotential operators are defined at each site, contrary to the link operator defined on the links in the Kogut-Susskind formalism. As demonstrated in the Fig. \ref{fig:pp}, the link operator on a particular link is reconstructed in terms of the prepotential operators residing at both of its ends. The constraint given in (\ref{eq:AGL_KS}) is translated to the `Abelian Gauss Law' constraint in prepotential formalism. This new constraint imposes that {\it the total number of prepotential operators at both ends of a particular link must be equal}. 

Having defined all the operators at each site that contains the Gauss law as well, one can now construct gauge invariant operators at each site. These local gauge invariant operators glued together with the `Abelian Gauss Law' constraint across neighboring sites yield all possible non-local Wilson loop operators of the theory. These local  loop operators acting on the local strong coupling vacuum (i.e no electric flux state) build the local loop basis of the theory. The local loop states at any lattice site, can be identified as a local snap shot of a Wilson loop state on that lattice. 

Going beyond SU(2), prepotential formalism exists for SU(3) \cite{Anishetty:2009ai,Anishetty:2009nh} as well as any SU(N) \cite{Raychowdhury:2013rwa,mathur2010n} gauge groups that share the same features.

\subsection{Loop-String-Hadron formalism}
\label{sec:lsh}
\noindent

Extending prepotential formalism beyond pure gauge theory leads to the loop-string-hadron (LSH) formalism of non-Abelian gauge theories. The complete LSH formalism for gauge group SU(2) is given in a recent work \cite{Raychowdhury:2019iki}. As mentioned before, the prepotential formalism comes with a local description of loop states at each lattice site for pure gauge theory along with the Abelian Gauss law constraints along with each link. Including staggered matter field to this framework is straightforward as both prepotentials and fermionic matter fields transform in the same way, i.e in the fundamental representation of the gauge group at each site. Now, in presence of matter, the complete set of local SU(2) invariant operators include the bilinears constructed out of (a) two components of matter (fermions), (b) one component of prepotential (bosons) and one component of matter (fermions), and (c) two components of prepotential (bosons) identified as local (a) hadrons, (b) strings and (c)loops of the theory.  The complete set of operators that form a closed algebra along with the Abelian Gauss law constraints discussed before leads to local snapshots of all possible non-local variables of the theory. The detail of this closed operator algebra along with explicit Hamiltonian written in LSH representation is given in \cite{Raychowdhury:2019iki}. 

\begin{figure}
\centering
 \includegraphics[width=1\textwidth]{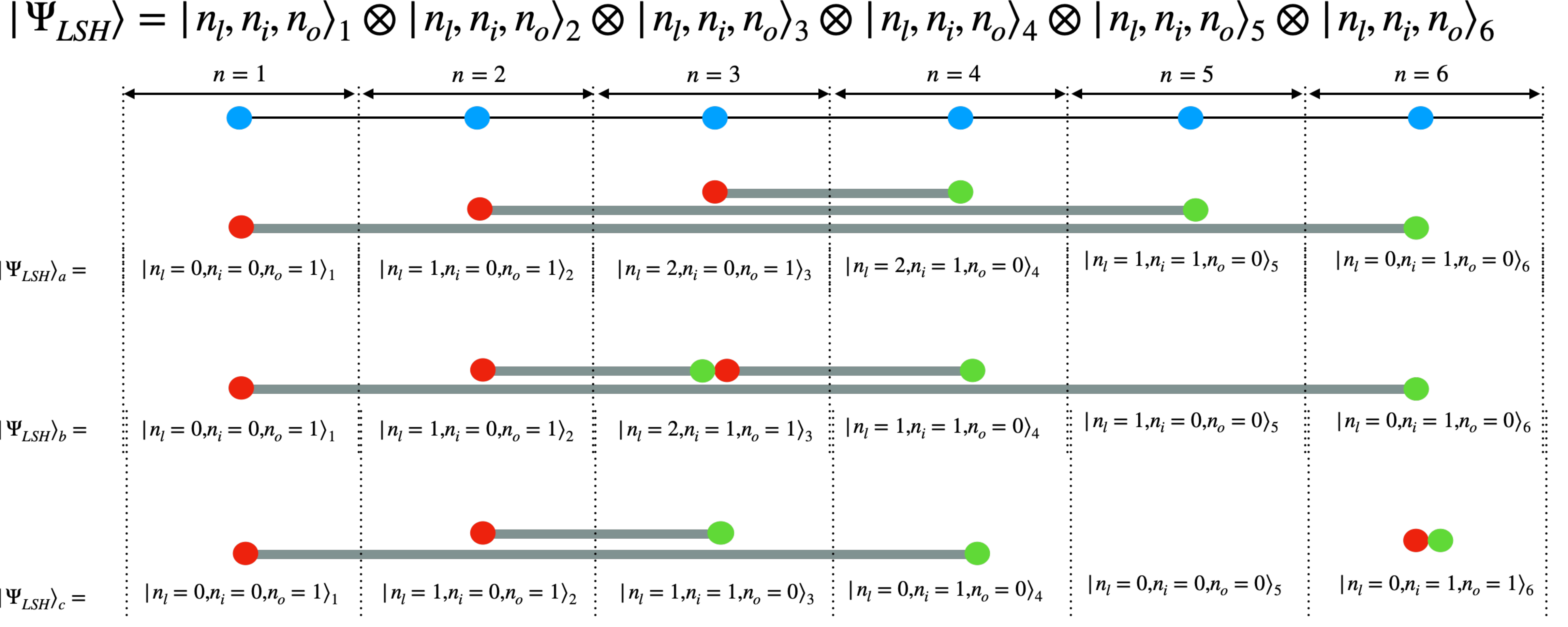}
\caption{Pictorial representation of gauge invariant non-local strings and hadrons on a $1$d spatial lattice characterized by local $n_l,n_i,n_o$ quantum numbers at each lattice site. Presence of a red(green) dot at one site denotes the outgoing(incoming) string quantum number to take non-zero value. When both of them are located at a single site, it denotes the presence of a hadron. Number of solid lines passing through a lattice site gives the value of loop quantum number at that site. The Abelian Gauss law constraint as given in (\ref{AGL:qq}) is satisfied by continuation of solid lines across each of the vertical dotted lines.}
\label{fig:lsh1d}       
\end{figure}

The simplest scenario of non-Abelian gauge theory coupled to matter field is in the $1+1$ dimension. The LSH basis in this case is described by three local quantum numbers $n_l, n_i, n_o$, that is the eigenbasis of the electric and mass term of the Hamiltonian.  In one spatial dimension, electric flux can only flow along one direction yielding one single loop quantum number $n_l$ which can take any integer value in the range $(0,\infty)$. However, depending on the matter content of each lattice site, a starting and/or ending point of a gauge invariant string state, i.e, electric flux lines connecting a matter-antimatter pair, can exist. This two end configuration of a string state is characterized by incoming and outgoing string quantum number $n_i, n_o$. The associated operator that creates a local string state acting on strong coupling vacuum consists of bilinears of a bosonic and a fermionic oscillator and hence the string quantum numbers can take values $(0,1)$ at each site. It is shown in \cite{Raychowdhury:2019iki} that the action of the local hadron operator is equivalent to action of both the incoming and outgoing string operator at one site, hence $n_l,n_i,n_o$ can completely characterize local LSH Hilbert space as illustrated in Fig. \ref{fig:lsh1d}. The Abelian Gauss law at neighboring lattice sites are given by:
\bea
\label{AGL:qq}
n_l(x)+n_o(x)\left[1-n_i(x)\right]=n_l(x+1)+n_i(x+1)\left[1-n_o(x+1)\right]
\eea

\begin{figure}
\centering
 \includegraphics[width=0.8\textwidth]{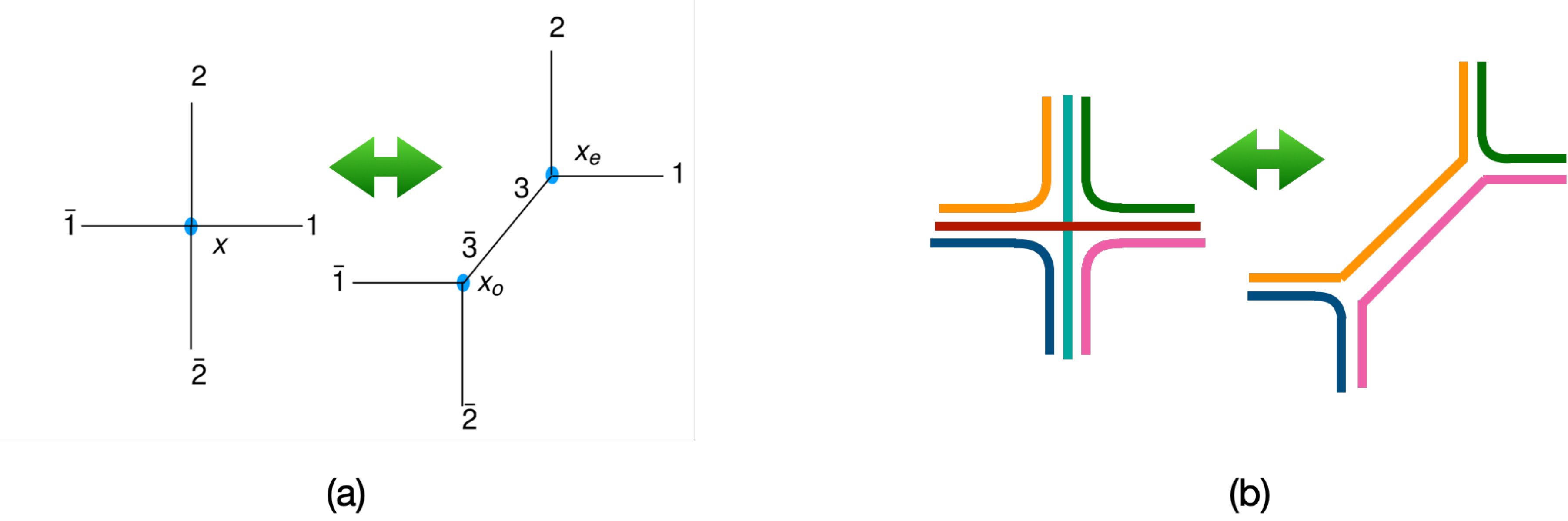}
\caption{Under the virtual point splitting scheme, a lattice site $x$ on a $2$-d lattice is split into two virtual sites $x_e,x_o$ separated by a virtual link along the direction $3-\bar 3$. The local loop configurations of the original lattice, i.e 6 possible orientation of loops are also split into local loops at each of the two new vertices. This allows three loop orientations at each site. So, the total number of local loop configurations at the original site equals to the total number of loop configurations at the equivalent two virtual sites. The continuity of flux lines along the virtual direction introduces one additional Abelian Gauss law on that site, which compensates the complicated Mandelstam constraint present in the original formalism. }
\label{fig-pointsplitting}       
\end{figure}
\begin{figure}
\centering
 \includegraphics[width=0.6\textwidth]{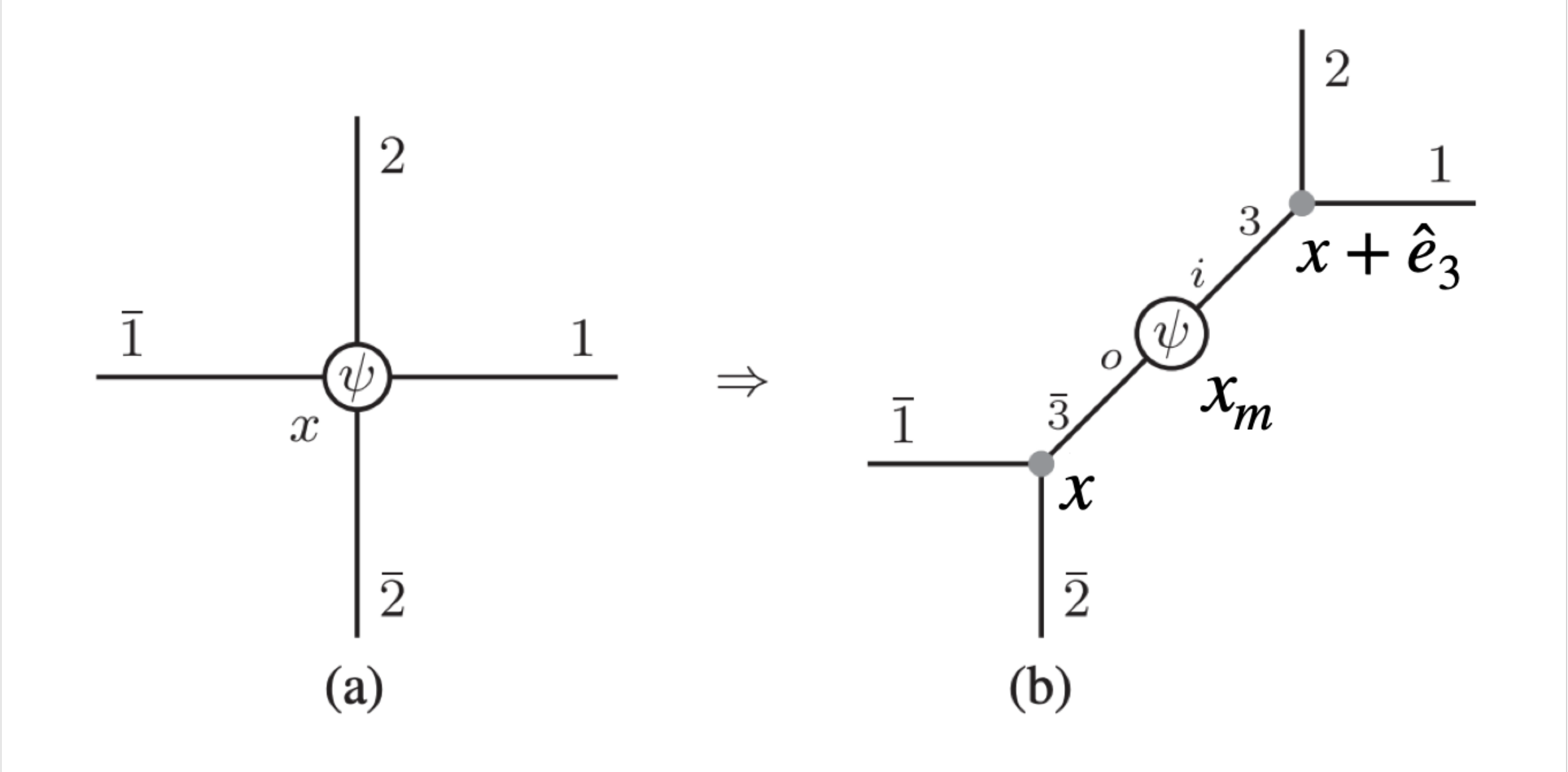}
\caption{For $d=2$, matter field residing at site x on the original lattice is mapped to the configuration of matter field residing on an additional site $x_m$ on the point split lattice. Locally $x_m$ looks like a site of $1$-d lattice connected to an incoming and outgoing link denoted by $i,o$ and carrying local LSH state $|n_l,n_i,n_o\rangle$.}
\label{fig:pointsplitmatter}       
\end{figure}

For pure gauge theory with spatial dimension two or more, there exists more than one local loop quantum number at each lattice site, forming an overcomplete local loop basis. The physical loop degrees of freedom can only be obtained after solving local version of Mandelstam constraint mentioned before and this task becomes more and more non-trivial with increasing spatial dimension. Following the trick of virtual point splitting prescribed in \cite{Raychowdhury:2018tfj,anishetty2018mass}, one could bypass solving the complicated and non-linear Mandelstam constraint at the cost of introducing virtual links and hence extra Abelian Gauss laws along the virtually split sites as illustrated in Fig. \ref{fig-pointsplitting} for two spatial dimension. This point splitting scheme can be generalized to any arbitrary spatial dimension. Now, at each site of the point split lattice one can define three local loop degrees of freedom and can characterize the local loop Hilbert space by three integer quantum numbers $l_{12}(x),l_{23}(x),l_{31}(x)$ taking integer values in the range $(0,\infty)$.
To include matter field at the original lattice site, one can introduce an additional virtual lattice site $x_m$ on the point split site as per Fig. \ref{fig:pointsplitmatter}. Locally, $x_m$ looks like a site of the $1-$d spatial lattice and contain local gauge invariant Hilbert space characterized by quantum numbers $n_l(x_m),n_i(x_m),n_o(x_m)$ as discussed before. Hence, within this scheme, coupling of matter to gauge field in any arbitrary dimension is as simple as the same in one spatial dimension.

The Abelian Gauss law along links are constrained to have equal number of prepotential or bosons at each end of the link. For $2$-d spatial lattice, these constraints are of the following form:
\bea
\mbox{Along direction $\hat e_1$: }~~~~~~~~~~~~~&& \nonumber \\ l_{12}(x)+l_{31}(x) &=& l_{12}(x+\hat e_1)+l_{31}(x+\hat e_1)  \\
\mbox{Along direction $\hat e_2$: }~~~~~~~~~~~~~&& \nonumber \\ l_{12}(x)+l_{23}(x) &=& l_{12}(x+\hat e_2)+l_{23}(x+\hat e_2)  \\
\mbox{Along virtual direction $\hat e_3$: }~~~~~~~~~~~&& \nonumber \\
l_{23}(x)+l_{31}(x) &=& n_l(x_m)+n_i(x_m)\left(1-n_o(x_m)\right) \\
n_l(x_m)+n_o(x_m)\left((1-n_i(x_m) \right)&=& l_{23}(x+\hat e_3)+l_{31}(x+\hat e_3) 
\eea

In summary, within the loop string hadron formalism of SU(2) lattice gauge theory, the SU(2) Gauss law is already solved and the only constraint that remain is Abelian Gauss law constraints, that is there for each link direction. These constraints are actually equivalent to Gauss law constraint present in Schwinger model, an U(1) gauge theory in 1+1 dimension. Hence, generalizing the strategies already developed for quantum simulation of Schwinger model should make significant progress towards quantum simulating  a non-Abelian gauge theory with gauge group SU(2). In the next section, we briefly mention some of the very recent progress in the context of quantum simulation that can utilize the LSH formalism and define state of the art of this research direction.

\section{Quantum Simulation}
\label{sec:app}
\noindent
As mentioned before, the progress towards quantum simulating even simplest non-Abelian gauge group is significantly less than the same for Schwinger model. There has been no analog experiment reported so far that can simulate a SU(2) gauge theory dynamics. 
There has been only one digital implementation of the rigid rotor representation of pure SU(2) gauge theory on a 1-dimensional ladder lattice consisting of few plaquettes for trotterization on IBM cloud quantum computer \cite{klco20202} and also analyzed numerically on a quantum annealer \cite{rahman20212}. However, this scheme is reported to be non-scalable for arbitrary lattice. 

 In a recent study \cite{new}, it has been established that the computational complexity of simulating LSH Hamiltonian in $1+1$ dimension is way less than other Hamiltonian formalisms of the same theory existing in the literature \cite{Zohar:2019ygc,Zohar:2018cwb,Kogut:1974ag,chandrasekharan1997quantum}. For gauge theory in one spatial dimension and with open boundary condition, one can always choose pure gauge that yields a purely fermionic formalism of the same theory without any Gauss law constraint and hence is easy for any practical computation. computational cost of simulating the purely fermionic Hamiltonian is comparable to that of LSH, but the fermionic formalism cannot be generalized to arbitrary dimension or arbitrary boundary condition. In this regard, LSH formalism seems more promising.

The purely fermionic formalism has been exploited in context of tensor network calculation of Hamiltonian simulation \cite{banuls2020review} and also for digital quantum simulation of SU(2) hadrons in a very recent work \cite{atas20212}. In the following two subsections we briefly mention some recent progress with both digital and analog quantum simulation of SU(2) gauge theory that has been made possible only by utilizing the LSH formalism discussed before and offers hope for much more progress in both of these directions in near future.

\subsection{Digital Quantum Computation}
\label{sec:digital}
\noindent
\subsubsection{Digitization:}
\begin{figure}
\centering
 \includegraphics[width=1\textwidth]{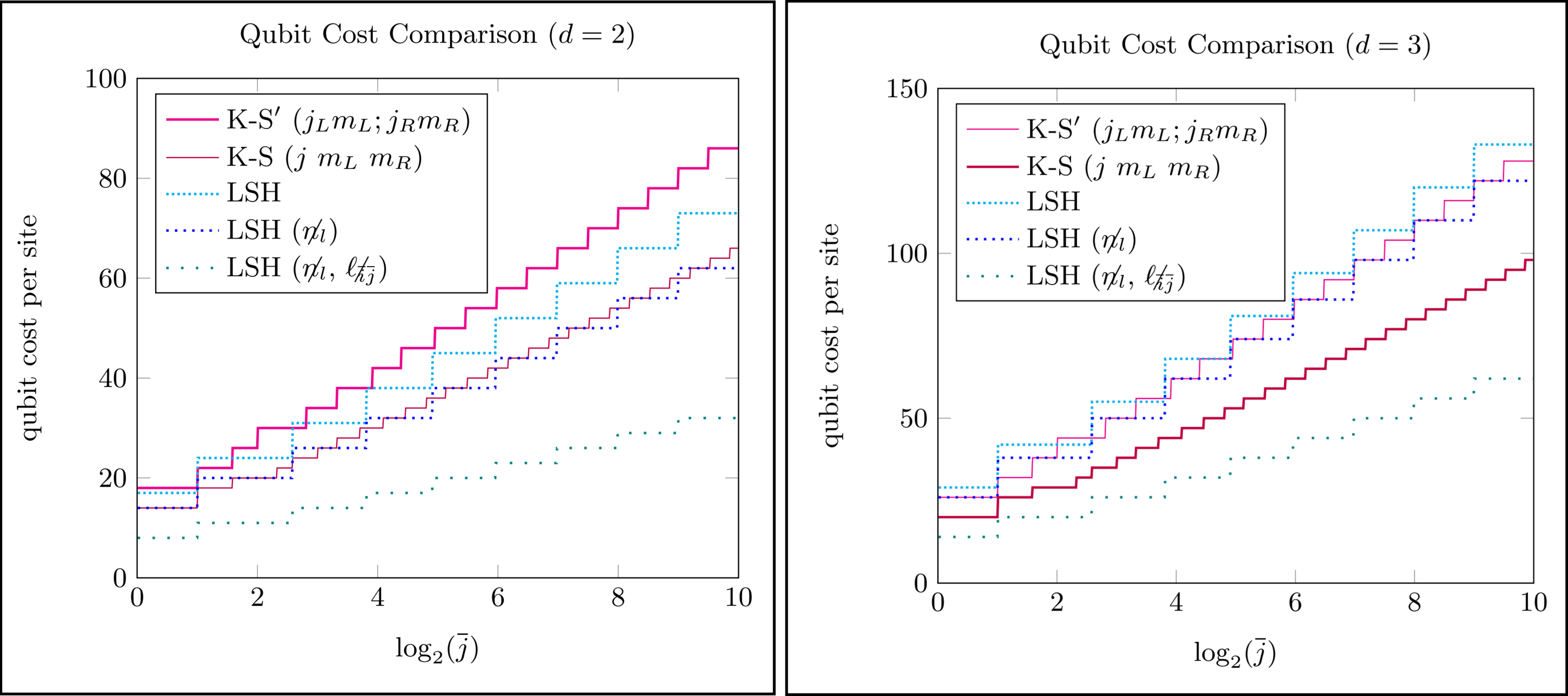}
\caption{Qubit cost per lattice site for (i) Kogut-Susskind formalism with local angular momentum basis without solving (\ref{eq:AGL_KS}); (ii) Kogut-Susskind formalism with rigit rotor basis; (iii) LSH formalism without solving any of the Abelian Gauss law constraints; (iv) LSH formalism after solving Abelian Gauss law constraint only along the link that contain matter (e.g. along direction $3-\bar 3$ in $d=2$); (v) LSH formalism after solving all of the Abelian Gauss law constraints. }
\label{fig:qubit_cost}       
\end{figure}

Digitization of LSH formalism is given in \cite{Raychowdhury:2018osk}. The string or fermionic quantum numbers are already binary in LSH formalism. The loop quantum numbers can be expanded in binary representation and hence can be represented by qubits in a straightforward way. Although the loop quantum numbers can take integer values in the range $(0,\infty)$, for practical purpose one must put an upper cut-off in these values. Note that, imposing the cut-off does not affect the symmetry of the theory and it still remains as SU(2) symmetries at each of the lattice sites. The cut-off actually acts on the irreps of SU(2), i.e instead of having infinite number of SU(2) irreps (characterized by $0\le j\le \infty$), one actually considers a finite set of irreps  (characterized by $0\le j\le j_{max}$) that still transform under the same SU(2). The SU(2) ladder operator (or, the flux-creation operator) is set to annihilate the highest allowed irrep to keep the dynamics of the theory confined within the cut-off. 
A similar cutoff is necessary to work with the conventional rigid rotor basis as well. For a given cutoff $j_{max}=\bar j$, the number of logical qubits required per lattice site can be computed as in \cite{Raychowdhury:2018osk}. A comparative analysis is quoted from \cite{Raychowdhury:2018osk} in Fig. \ref{fig:qubit_cost}. This analysis shows the efficiency of LSH formalism in terms of qubit-cost per site when all (or even some of) the Abelian Gauss law constraints are solved. 

\subsubsection{Physical Hilbert space:}
The LSH states by construction satisfy the SU(2) Gauss' law. As discussed in detail in the last section, the local LSH states at neighboring lattice sites must solve Abelian Gauss law constraints along each link to yield the non-local Wilson loop and string configurations of the theory. Note that, the Hamiltonian commutes with the Abelian Gauss law constraint. So, in principle if one can prepare an initial state on the lattice that satisfies the constraint everywhere on the lattice, i.e a physical LSH state, it should evolve to some other LSH states under this Hamiltonian. However, in the noisy hardwares of the NISQ era, one cannot expect such an ideal evolution. Rather, even if one starts with a physical LSH state, due to errors of the hardware there will be a large probability to end up with an unphysical LSH state, that would not satisfy the Abelian Gauss law everywhere on the lattice. At this point, \cite{Raychowdhury:2018osk} provides an oracle (or digital quantum circuit) to check for the physicality criterion of LSH states. This physicality oracle is expected to act as an useful error mitigation tool for digital quantum simulation of SU(2) gauge theory. 

A similar physicality oracle had also been constructed for U(1) gauge theory \cite{stryker2019oracles} that checks if a state satisfies U(1) Gauss law constraint at a lattice site for the same purpose. However, the striking feature of LSH formalism being as simple as the Schwinger model is reflected in the fact that the SU(2) physicality oracle, constructed for any arbitrary dimension is only as complex as the physicality oracle of U(1) gauge theory in one spatial dimension. 

\subsubsection{Hamiltonian evolution:} 
Hamiltonian evolution of a physical state within the rigit rotor basis of Kogut-susskind formalism is quite nontrivial to realize on a digital quantum computer since, the physical states are linear combination of many angular momentum states (and this number of terms in each linear combination scales exponentially). To time evolve a physical state, one would require to map each angular momentum state to different qubits and let all of those evolve simultaneously that is practically impossible at least with the NISQ era hardware. Whereas, the LSH basis is already gauge invariant and is one-sparse, i.e the Hamiltonian acting on one particular LSH state would give another LSH state (not a linear combination). One should be able to utilize this feature to obtain quantum algorithm for time evolution of the theory like the algorithm already developed for Schwinger model in \cite{shaw2020quantum}. Work is in progress along this direction. 

\subsection{Analog Quantum Simulation}
\label{sec:analog}
\noindent

The  concept  of  analog  quantum  simulation  involves mimicking  a  quantum  system  described  by  a  Hamiltonian (simulated Hamiltonian) by another completely different quantum system described by some other Hamiltonian (simulating Hamiltonian). The idea is, if it is hard to analyze the first Hamiltonian mathematically or numerically, it can be mapped to the second Hamiltonian,and that Hamiltonian can be studied in an experiment by suitably tuning the parameters.  Systems of ultracold atoms \cite{bloch2012quantum} or ions \cite{blatt2012quantum} trapped in optical lattices serve as excellent quantum simulators, as the relevant parameters can  be  precisely  measured  and  controlled. 
As mentioned in the introduction, there has been several proposals for simulating both Abelian and non-Abelian gauge theories in an analog experiment around the first half of the last decade. Starting from the second half of the last decade, there has been quite a few experimental demonstration of analog simulation of the Abelian theory for small lattice or even some scalable building block of the theory. However, any experimental demonstration of simulating simple non-Abelian toy model has not been reported so far. 
There is many fold complication associated with analog quantum simulation of SU(2) gauge theory, including the non-triviality in realizing the notion of gauge invariance and also engineering the Hamiltonian operator as some experimentally feasible many body interaction in the simulating system. 

\begin{figure}
\centering
 \includegraphics[width=1\textwidth]{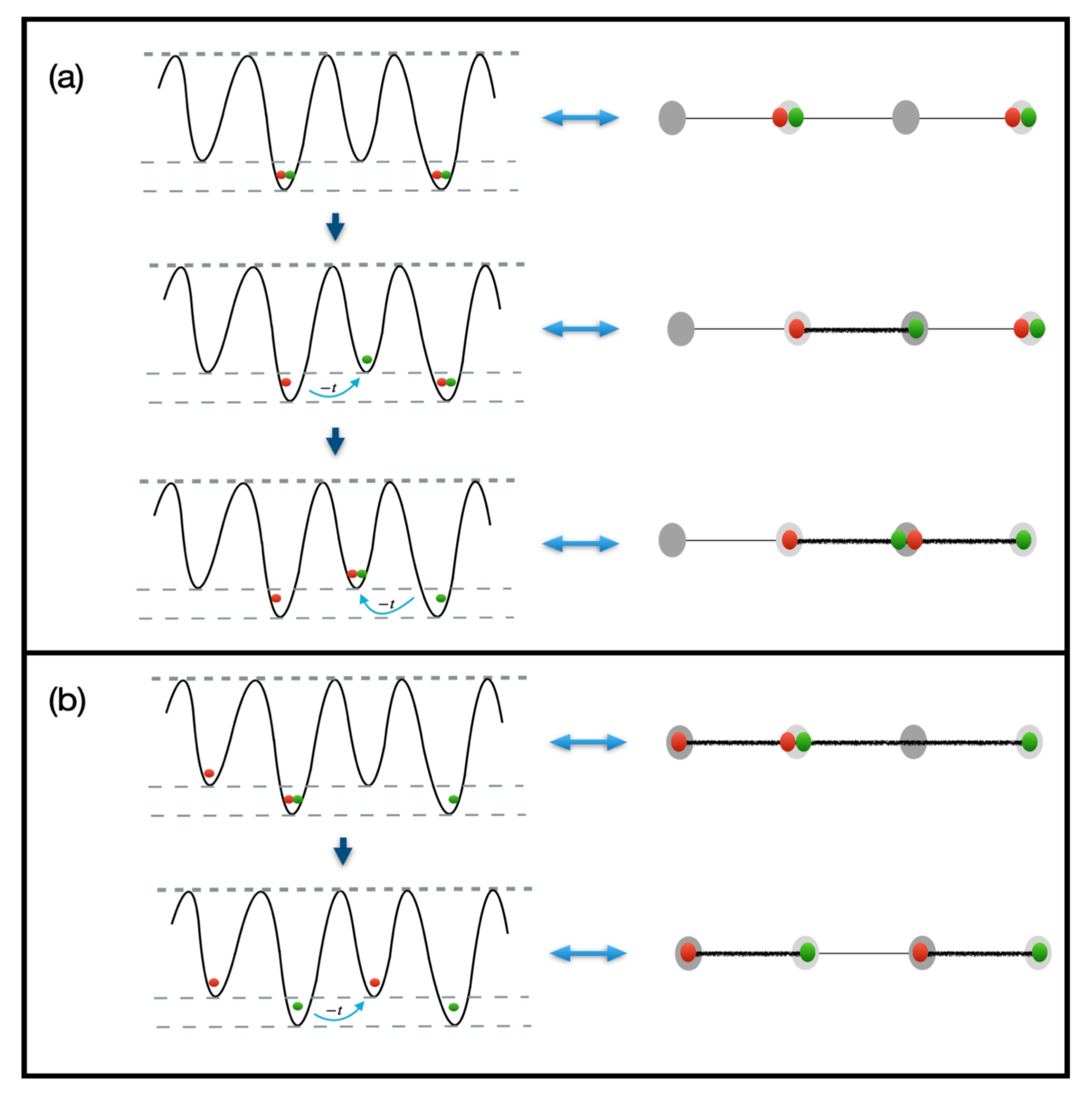}
\caption{Cartoon representation of dynamics of 1d ionic Hubbard model mimicking that of SU(2) lattice gauge theory in one spatial dimension. (a) Initial state: fully filled odd sites and empty even sites mimicking the strong coupling vacuum consisting of no particles ($n_i=0,n_o=0$ on even sites) and no antiparticles ($n_i=1,n_o=1$ on odd sites). Under Hamiltonian evolution, one atom hops from an odd site to neighboring even site in Hubbard model, that mimics creation of a particle antiparticle pair at two neighboring staggered site of the gauge theory, connected by one unit of flux to form a gauge singlet string configuration. One further hopping as shown in the figure mimics the dynamics in gauge theory as elongation of the string and creation a baryon on one site. In these three states, the total number of particle (antiparticle) for the gauge theory are respectively $0,1,2$. (b) Ionic Hubbard model dynamics is mimicking string breaking dynamics of gauge theory. Starting from a string of length 3 unit, pair production occurs and the initial string breaks into two smaller strings.}
\label{fig:cartoon}       
\end{figure}

In a very recent proposal \cite{dasgupta2020cold}, an analog quantum simulator to simulate SU(2) lattice gauge theory dynamics in $1+1$ dimension has been proposed that is well within the reach of present day's cold atom experiment. This proposal is based upon the LSH formalism and hence manifestly SU(2) gauge invariant. The starting point of this proposal is a mean field approximation of the loop configuration and that was valid in the weak coupling regime. However, the novel feature of this proposal is obtained by fine tuning the the atomic interaction to compensate for the error made due to the mean field approximation. As a result, with suitable tuning of parameters the simulated dynamics do mimic the dynamics of the full Kogut-Susskind theory for a wide range of coupling parameters, starting from  the very weak coupling to intermediate coupling regime. \cite{dasgupta2020cold} also provide numerical benchmarks for the simulating and simulated dynamics using the parameter values of the simulating system that has already been achieved in an experiment \cite{messer,scherg}. The proposed simulating system consists of a set of fermionic atoms trapped into bipartite optical lattice and governed by ionic Hubbard model Hamiltonian. The cartoon of the simulating and simulated dynamics is described in Fig. \ref{fig:cartoon}.
This scheme can simulate the dynamics accurately for a large lattice that is beyond the scope of classical computation.

This particular proposal is restricted to one spatial dimension and for intermediate to weak coupling regime. Work is in progress to construct quantum simulators for $2+1$ dimensional SU(2) lattice gauge theory.

\section{Summary and future directions}
\label{sec:summary}
\noindent

In this review, we briefly mentioned the challenges to proceed with quantum computing lattice QCD in the upcoming quantum computer era. As the very first step, we discuss some recent progress towards quantum simulation SU(2) lattice gauge theory in both digital and analog way. 

Analyzing all available Hamiltonian formulations of Yang Mills theory on lattice, one particular formalism namely the loop string hadron formalism has found to be convenient for quantum computing SU(2) gauge theory. Recent state of the art developments of both digital and analog simulation of SU(2) gauge theory based on this novel LSH framework is also discussed. Many research groups throughout the globe is presently working with SU(2) gauge theory from quantum information science perspective. The rapidly growing community of researchers with this particular research interest is expected to contribute much more in this field in next few years. Even before adequate quantum resource becomes available, tensor network calculations performed for SU(2) lattice gauge theory on a large lattice would give important insights. Work is in progress along this direction too. 

Towards the final goal of quantum simulating QCD, there has been even less progress, namely \cite{ciavarella2021trailhead} gives the very first result of digital quantum computation of two plaquette SU(3) gauge theory. prepotential formalism for gauge group SU(3) \cite{Anishetty:2009nh} already exists in literature. Refinement of the same to develop LSH formalism for SU(3) gauge theory will no doubt be a significant  step towards the final goal.

\section*{Acknowledgement} IR would like to thank Pushan Majumdar for numerous discussions, his encouragement and enthusiasm towards this emerging field of quantum information science to be applied for lattice gauge theories. IR would also like to mention her sincere gratitude to Pushan Majumdar for his constant support during the difficult period faced by her after having a career break and helping her to return to research career.

IR would like thank collabrators Manu Mathur, Ramesh Anishetty, Jesse Stryker, Zohreh Davoudi, Raka Dasgupta for fruitful collaborations at different stages of development towards this research goal. IR would also like to thank David B. Kaplan for direction and support towards this particular research direction. IR is supported by the U.S. Department of Energy (DOE),Office of Science, Office of Advanced Scientific Computing Research (ASCR) Quantum Computing Application Teams (QCAT) program, under fieldwork Proposal No.ERKJ347.

\bibliographystyle{spphys}       
\bibliography{bibi.bib}   
\end{document}